\documentclass[12pt]{article}

\usepackage{epsf}
\usepackage{amssymb}
\usepackage{amsfonts}
\def\r{{\cal R}}
\title{$f(\r)$ Quantum Cosmology}
\author{Ali Shojai, and  Fatimah Shojai\\
Department of Physics, University of Tehran, Tehran, Iran.} 
\date{}
\begin{document}
\maketitle
\begin{abstract}
We have quantized a flat cosmological model in the context of the metric $f(\r)$ models, using the causal Bohmian quantum theory. The equations are solved and then we have obtained how the quantum corrections influence the classical equations.
\end{abstract}
\section{Introduction}
Extension of the Einstein--Hilbert Lagrangian from $\sqrt{-g}\r$ to the general form $\sqrt{-g}f(\r)$ is discussed in the literature\cite{sot1} and used to answer a wide range of questions in general relativity such as the dynamics of galaxies and the present accelerated expansion of the universe\cite{carr1}.

Corrections to the standard general relativity results may be investigated in two basic areas, the spherically symmetric solution\cite{clas} and in the cosmological models\cite{carr1}. These are basically done in the classical regimes, although there are some works\cite{quan} dealing with quantum models.

Investigation of the problem how and how much the quantum effects gives additional corrections to the standard general relativity seems to be an important task at least in domains in which one expects large quantum corrections.

Although this can be done in the standard framework of quantum mechanics, as it is discussed in the literature\cite{quan}, using the causal interpretation of quantum mechanics introduced by de-Broglie and Bohm\cite{bohm} is a good idea at least for gravity and cosmology. This is because Bohmian quantum mechanics does not suffer from some essential conceptual problems of the standard quantum mechanics which show themselves more transparently in gravity. In addition Bohmian quantum gravity presents a framework which is simply be extended both in gravity side and in quantum mechanical side\cite{self}.

In this paper, we shall investigate the quantum corrections to a flat cosmological model in $f(\r)$ gravity using Bohmian quantum mechanics. In the next section we briefly review the de-Broglie--Bohm method of causal interpretation of quantum mechanics. Next we shall use an effective lagrangian recently proposed in \cite {souza} and quantize it according to the de-Broglie--Bohm approach. Then we shall apply it to some specific functions $f$ and discuss how the quantum effects change the classical equations and the physical quantities.
\section{The Method of Causal Interpretation}
Here we present a very brief review of causal quantum mechanics. For a more detailed review the reader can refer to \cite{bohm}. 

Bohmian quantum mechanics is motivated by writing $\psi=Re^{iS/\hbar}$ in the non-relativistic quantum mechanics. The Shr\"odinger equation then gives the following two equations:
\begin{equation}
\frac{\partial R^2}{\partial t}+\vec{\nabla}\cdot\left (R^2\frac{\vec{\nabla}S}{m}\right )=0
\end{equation}
and
\begin{equation}
\frac{\left | \vec{\nabla}S\right |^2}{2m}+V+{\cal Q}=-\frac{\partial S}{\partial t}
\end{equation}
in which the quantum potential is given by:
\begin{equation}
{\cal Q}=-\frac{\hbar^2}{2m}\frac{\nabla^2R}{R}
\end{equation}
Bringing in mind that $R^2$ gives the ensemble density of the particle under consideration (Born statistical postulate) these resemble the Hamilton--Jacobi theory of a system with the quantum Hamiltonian ${\cal H}_{\cal Q}={\cal H}_c+{\cal Q}$ where ${\cal H}_c$ is the classical Hamiltonian. It is shown\cite{bohm} that particle trajectories arising from this quantum Hamiltonian reproduce all quantum mechanical results including the measurement principle.

Accordingly one can generalize the above ideas and introduces the following way for quantizing any system, \textit{the method of causal interpretation}:
\begin{enumerate}
\item Any quantum system is defined by its trajectory $q(t)$ and a self-field $\psi(q)$ (the wave function).
\item The self-field satisfy an appropriate wave equation, derived from the Dirac's quantization scheme (this is the Schr\"odinger equation for a non-relativistic particle).
\item System trajectory is obtained by the quantum Hamiltonian ${\cal H}_{\cal Q}={\cal H}_c+{\cal Q}$ in which ${\cal Q}$ is defined in terms of the self-field $\psi$. The form of the quantum potential is obtained by setting $\psi=Re^{iS/\hbar}$ in the wave equation.
\end{enumerate}
\section{Classical Model}
The action functional of $f(\r)$ gravity is:
\begin{equation}
{\cal A}=\frac{1}{2}\int d^4x \sqrt{-g}f(\r)+{\cal A}_{\textrm{matter}}
\end{equation}
There are two approaches for deriving the equations of motion from this action. First one regards the metric as the dynamical variable and use the Christoffel symbols as the connections need to define $\r$. Second approach is the metric--affine approach in which the metric and the connection are assumed independent dynamical variables. For the Einstein--Hilbert action for which $f(\r)=\r$, the field equations of the second approach leads to the equality of the connection and the Christoffel symbols, provided that the matter action do not depends upon the connection. Therefore the two approaches are identical in this case\cite{wald}. But for a general $f(\r)$ theory the two approaches are not necessarily equivalent except in the absence of matter in which case one gets Einstein's field equation with an arbitrary cosmological constant\cite{ferr}. 

Here we are working with metric approach. The equations of motion are then:
\begin{equation}
f'\r_{\mu\nu}-\frac{1}{2}f g_{\mu\nu}=\nabla_\mu\nabla_\nu f' -g_{\mu\nu}\Box f' +{\cal T}_{\mu\nu}
\end{equation}
where a prime denotes differentiation with respect to $\r$.

In this paper we are interested in the $f(\r)$ cosmological models. In order to make things simple, let us consider the case of a flat FLRW metric\cite{souza}:
\begin{equation}
ds^2=-dt^2+a^2(t)(dx^2+dy^2+dz^2)
\label{eq1}
\end{equation}
The above field equations in vacuum simplify to:
\begin{equation}
H^2=\frac{1}{3f'}\left [ \frac{1}{2}(\r f'-f)-3H\dot{\r}f''\right ]
\label{eq2}
\end{equation}
and
\begin{equation}
2\dot{H}+3H^2=-\frac{1}{f'}\left [ f'''\dot{\r}^2+2H\dot{\r}f''+\ddot{\r}f'' +\frac{1}{2}(f-\r f') \right ]
\label{eq3}
\end{equation}
in which $H$ is the Hubble parameter:
\begin{equation}
H=\frac{\dot{a}}{a}
\end{equation}
and a dot over any quantity represents differentiation with respect to time.

It is important to note that the equations of motion are expressed solely in terms of $H$ and $\r$ and their time derivatives. This is a special case for flat model. In the case of non flat universe the equations of motion also contain terms expressed explicitly in terms of $a$. Taking $a$ and $\r$ as the dynamical variables, an effective Lagrangian can be introduced as\cite{souza}:
\begin{equation}
{\cal L}_{\textrm{eff.}}={\cal L}_{\textrm{eff.}}(a,\dot{a},\r,\dot{\r})=a^3\left [ 6H^2f'+6Hf''\dot{\r}+f'\r-f\right ]
\label{lag}
\end{equation}
The canonical momenta are then:
\begin{equation}
p_a=12a\dot{a}f'+6a^2f''\dot{\r}
\end{equation}
and
\begin{equation}
p_\r=6a^2\dot{a}f''
\end{equation}
and the Hamiltonian is:
\begin{equation}
{\cal H}=\frac{p_ap_\r}{6a^2f''}-\frac{f' p_\r^2}{6a^3f''^2}-a^3f'\r+a^3f
\end{equation}
The Hamilton equations:
\begin{equation}
\{a,{\cal H}\}=\dot{a}
\end{equation}
\begin{equation}
\{\r,{\cal H}\}=\dot{\r}
\end{equation}
\begin{equation}
\{p_a,{\cal H}\}=\dot{p_a}
\end{equation}
\begin{equation}
\{p_\r,{\cal H}\}=\dot{p_\r}
\end{equation}
leads to equation (\ref{eq2}) and the relation:
\begin{equation}
\r=6\left ( \frac{\ddot{a}}{a}+\frac{\dot{a}^2}{a^2} \right )
\end{equation}
The equation (\ref{eq3}) is identical with
\begin{equation}
{\cal H}=0
\end{equation}
which represents the time reparametrization invariance of the system.

Alternatively, one can use the Hamilton--Jacobi equation which is identical to the equations of motion:
\begin{equation}
\frac{1}{6a^2f''}\frac{\partial S}{\partial a}\frac{\partial S}{\partial \r} -\frac{f'}{6a^3f''^2}\left ( \frac{\partial S}{\partial \r}\right )^2-a^3f'\r+a^3f=0
\end{equation}

At this end, it must be noted that as it is shown in \cite{souza} the equations of motion have the fixed point $\dot{H}=\dot{R}=0$ which is de-Sitter space--time.
\section{Quantum Model}
The quantum version of the model of the previous section can be achieved via the method of canonical quantization. The Hamiltonian  constraint is now:
\begin{equation}
-\hbar^2\frac{\partial^2\psi}{\partial\r^2}+\hbar^2a g(\r)\frac{\partial^2\psi}{\partial\r\partial a}+a^6 U(\r)\psi=0
\end{equation}
in which
\begin{equation}
g=\frac{f''}{f'}
\end{equation}
\begin{equation}
U=6f''^2\left ( \r-\frac{f}{f'} \right )
\end{equation}
In order to obtain the Bohmian trajectories we set:
\begin{equation}
\psi=\Omega e^{iS/\hbar}
\end{equation}
and substitute this into wave equation. This leads to the continuity equation
\begin{equation}
\frac{\partial}{\partial \r}\left ( \Omega^2 \frac{\partial S}{\partial \r} \right ) -\frac{ag}{2}\frac{\partial}{\partial a}\left ( \Omega^2 \frac{\partial S}{\partial \r} \right ) -\frac{ag}{2}\frac{\partial}{\partial \r}\left ( \Omega^2 \frac{\partial S}{\partial a} \right )=0
\end{equation}
and the modified Hamilton--Jacobi equation
\begin{equation}
{\cal H}(a,\r,p_a=\partial S/\partial a,p_\r=\partial S/\partial \r)+{\cal Q}=0
\label{QHJ}
\end{equation}
in which the quantum potential is defined as:
\begin{equation}
{\cal Q}=\frac{\hbar^2f'}{6a^3f''^2}\left ( \frac{\Omega''}{\Omega}-a\frac{f''}{f'}\frac{\Omega'^*}{\Omega} \right )
\end{equation}
where a $*$ superscript represents differentiation with respect to $a$.

This shows that the quantum Hamiltonian is given by:
\begin{equation}
{\cal H}_{\cal Q}={\cal H}+{\cal Q}
\label{HQ}
\end{equation}
and the quantum equations of motion are:
\begin{equation}
\{a,{\cal H}_{\cal Q}\}=\dot{a}
\end{equation}
\begin{equation}
\{\r,{\cal H}_{\cal Q}\}=\dot{\r}
\end{equation}
\begin{equation}
\{p_a,{\cal H}_{\cal Q}\}=\dot{p_a}
\end{equation}
\begin{equation}
\{p_\r,{\cal H}_{\cal Q}\}=\dot{p_\r}
\end{equation}
leading to:
\begin{equation}
p_\r=6a^2\dot{a}f''
\label{e1}
\end{equation}
\begin{equation}
p_a=12a\dot{a}f'+6a^2\dot{\r}f''
\label{e2}
\end{equation}
\begin{equation}
2\dot{H}+3H^2+\frac{1}{f'}\left [ f'''\dot{\r}^2+2H\dot{\r}f''+\ddot{\r}f'' +\frac{1}{2}(f-\r f') \right ]=-\frac{1}{6a^2f'}\frac{\partial{\cal Q}}{\partial a}
\label{qeq2}
\end{equation}
\begin{equation}
\r=6\left (\frac{\ddot{a}}{a}+\frac{\dot{a}^2}{a^2}\right ) +\frac{1}{a^3f''}\frac{\partial{\cal Q}}{\partial\r}
\label{qeq3}
\end{equation}

And the quantum Hamiltonian constraint ${\cal H}_{\cal Q}=0$ leads to:
\begin{equation}
H^2=\frac{1}{3f'}\left [ \frac{1}{2}(\r f'-f)-3H\dot{\r}f''-\frac{{\cal Q}}{2a^3}\right ]
\label{qeq1}
\end{equation}
The fact that the Hamilton equations derived from the Bohmian quantum Hamiltonian (\ref{HQ}) are identical with the Hamilton--Jacobi equation (\ref{QHJ}) is not trivial, at least for geometrodynamics. It is shown that for Bohmian quantum geometrodynamics (i.e. not only for the cosmological minisuperspace, but also for any general superspace) the two are identical and that the Bohmian evolution is consistent with the quantum Hamiltonian and 3-diffeomorphism constraints\cite{pin,khod}.

The fixed point of the equations of motion can be derived via letting $\dot{H}=\dot{\r}=0$ leading to:
\begin{equation}
\r=12H^2+\frac{1}{a^3f''}\frac{\partial {\cal Q}}{\partial \r}
\end{equation}
\begin{equation}
3H^2=\frac{\r f'-f}{2f'}-\frac{{\cal Q}}{2a^3f'}
\end{equation}
\begin{equation}
{\cal Q}=\frac{a}{3}\frac{\partial{\cal Q}}{\partial a}
\label{sta}
\end{equation}
The last equation shows that not for any $f(\r)$ and any solution a fixed point exists. Only when ${\cal Q}=G(\r)a^3$ where $G$ is a general function, we have an exact fixed point, otherwise in regimes where this last equation holds approximately, we have an approximate de-Sitter space--time. Whether this fixed point is a stable solution or not highly depends on the quantum state of the universe.

It is possible to define an effective equation of state for the model via identifying the right hand side of equation (\ref{qeq1}) with effective density and the right hand side of equation (\ref{qeq2}) with effective pressure and treating $1/f'$ as the effective gravitational coupling constant. We have:
\begin{equation}
\rho_{eff}=\frac{\r f'-f}{2}-3H\dot{\r}f''-\frac{{\cal Q}}{2a^3}
\end{equation}
\begin{equation}
p_{eff}=\dot{\r}^2f'''+2H\dot{\r}f''+\ddot{\r}f''+\frac{f-\r f'}{2}+\frac{1}{6a^2}\frac{\partial{\cal Q}}{\partial a}
\end{equation}
\begin{equation}
w_{eff}=\frac{p_{eff}}{\rho_{eff}}=-1-\frac{2}{3}\frac{\dot{H}}{H^2}
\end{equation}
For the fixed point we have $w_{eff}=-1$ in accordance with type Ia supernovae results.

It is obvious that the fact that above equations can lead to accelerated expansion has its origin both in $f(\r)$ theory and in quantum effects. That is to say without postulation of dark energy, effects of $f(\r)$ and quantum contributions can accelerate the expansion of the universe. Whether which effect is dominant depends on the form of $f$ and the quantum state of the system. 
\section{Solutions and Results}
Usually $f(\r)$ is equal to $\r$ plus some additional terms. In order to have dimension of curvature for $f$, this additional terms have to contain a parameter having dimension of curvature. Therefore generally one can write:
\begin{equation}
f=\r+\beta \r_0 z(\r/\r_0)
\end{equation}
in which $\beta$ is a dimensionless parameter, $\r_0$ is a constant of dimension of curvature and $z$ is an arbitrary function. We shall assume that $\beta$ is a small parameter. Expanding the wave function in terms of the powers of $\beta$:
\begin{equation}
\psi=\psi_0+\beta\psi_1+\beta^2\psi_2+\cdots
\end{equation}
the wave equation reduces to the following at each order of expansion:
\begin{itemize}
\item{Zeroth order:}
\begin{equation}
\psi_0''=0
\end{equation}
\item{First order:}
\begin{equation}
\psi_1''+2\r_0 z'\psi_0''-a\r_0 z''\psi_0'^*=0
\end{equation}
\item{Second order:}
\begin{equation}
\psi_2''+\r_0 z'\psi_1''-a\r_0 z''\psi_1'^*=0
\end{equation}
\item{Third order:}
\begin{equation}
\psi_3''+\r_0 z'\psi_2''-a\r_0 z''\psi_2'^*-\frac{6a^6}{\hbar^2}\r_0^3 z''^2(\r z' -z)\psi_0=0
\end{equation}
\end{itemize}
The solution up to third order is then:
\[
\psi=A_1(a)+A_2(a)\r+\beta \left( A_3(a)+A_4(a)\r+a\r_0 zA_2^*(a)\right ) +
\]
\[
\beta^2 \left ( A_5(a)+ A_6(a)\r + a\r_0 zA_4^*(a)+\frac{1}{2}A_2^{**}(a)\int d\r \r_0^2 z'^2 \right ) +
\]
\[
\beta^3\left ( \frac{6a^6A_1(a)}{\hbar^2}\int d\r\int d\r \r_0^3 z''^2(\r z'-z)+\right .
\]
\[
\frac{6a^6A_2(a)}{\hbar^2}\int d\r\int d\r \r_0^3 z''^2(\r z'-z)\r+\frac{a^3A_2^{***}(a)}{6}\int d\r \r_0^3 z'^3 +
\]
\begin{equation}
\left . \frac{a^2A_4^{**}(a)}{2}\int d\r \r_0^2 z'^2+aA_6^{*}(a)\r_0 z+A_7(a)\r +A_8(a) \right ) +\cdots
\end{equation}
in which $A_1(a)\cdots A_8(a)$ are arbitrary functions of the scale factor arising as integration factors.

In order to extract physical results from this solution, we make some simplifications. We assume $A_2=A_3=A_4=A_5=A_6=A_7=0$ without losing general character of this wave function. In addition we set $A_1=1$ and $A_8=i$.
To justify this point it should be noted that for a real wavefunction which we name it \textit{pure quantum state}\cite{pure} the canonical momenta are zero. It is shown that \cite{pin}, in the case of gravity this corresponds to the strong gravity limit of general relativity which is governed by the Carroll group. This choice of $A_1$ and $A_8$, leads to a complex wavefunction and thus non zero $p_a$ and $p_\r$ which according to equations (\ref{e1}) and (\ref{e2}) corresponds to a time varying scale factor.

Therefore a simple solution up to the third order is:
\begin{equation}
1+\beta^3 i+\frac{\beta^3}{\hbar^2}a^6 p(\r)
\end{equation}
in which
\begin{equation}
p(\r)=\int d\r \int d\r 6\r_0^3 z''(\r z'-z)
\end{equation}
The quantum potential can be computed as:
\begin{equation}
{\cal Q}=\frac{\hbar^2}{6\beta^2a^3\r_0^2 z''^2}\left [ \frac{p''}{p}+\beta \r_0 z'\frac{p''}{p}-6\beta \r_0 z''\frac{p'}{p}\right ]
\end{equation}
The appearance of the negative powers of $\beta$ is not an alarm, because one never can put it equal to zero in this model as the lagrangian (\ref{lag}) does not give the correct equations of motion for the case $f=\r$.

Using the relation (\ref{qeq3}) the difference of the quantum curvature scalar and classical curvature scalar can be defined as:
\begin{equation}
\r -\r_c\equiv \frac{1}{a^3f''}\frac{\partial{\cal Q}}{\partial\r}=F(a,\r)
\end{equation}
This gives the quantum correction to the curvature scalar as a function of the scale factor and curvature scalar. 

\begin{figure}
\begin{center}
\epsffile{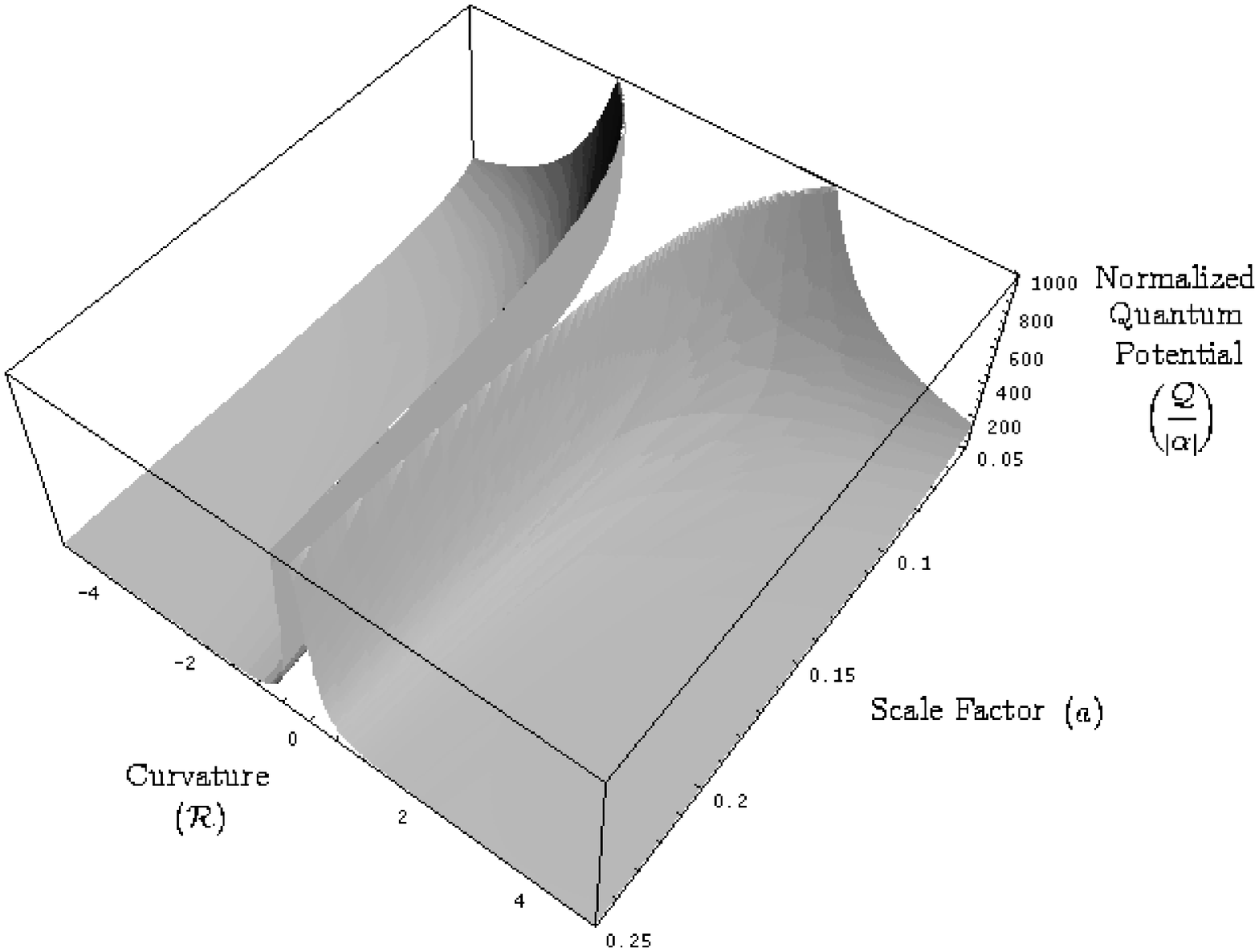}
\end{center}
\caption{Normalized quantum potential (${\cal Q}/|\alpha|$) as a function of the scale factor ($a$) and quantum curvature ($\r$) for the case $f(\r)=\r+\alpha\r^2$. $\alpha$ is chosen to be $|\alpha|^3=\hbar^2=\ell_p^4$.}
\label{fig1}
\end{figure}

\begin{figure}
\begin{center}
\epsffile{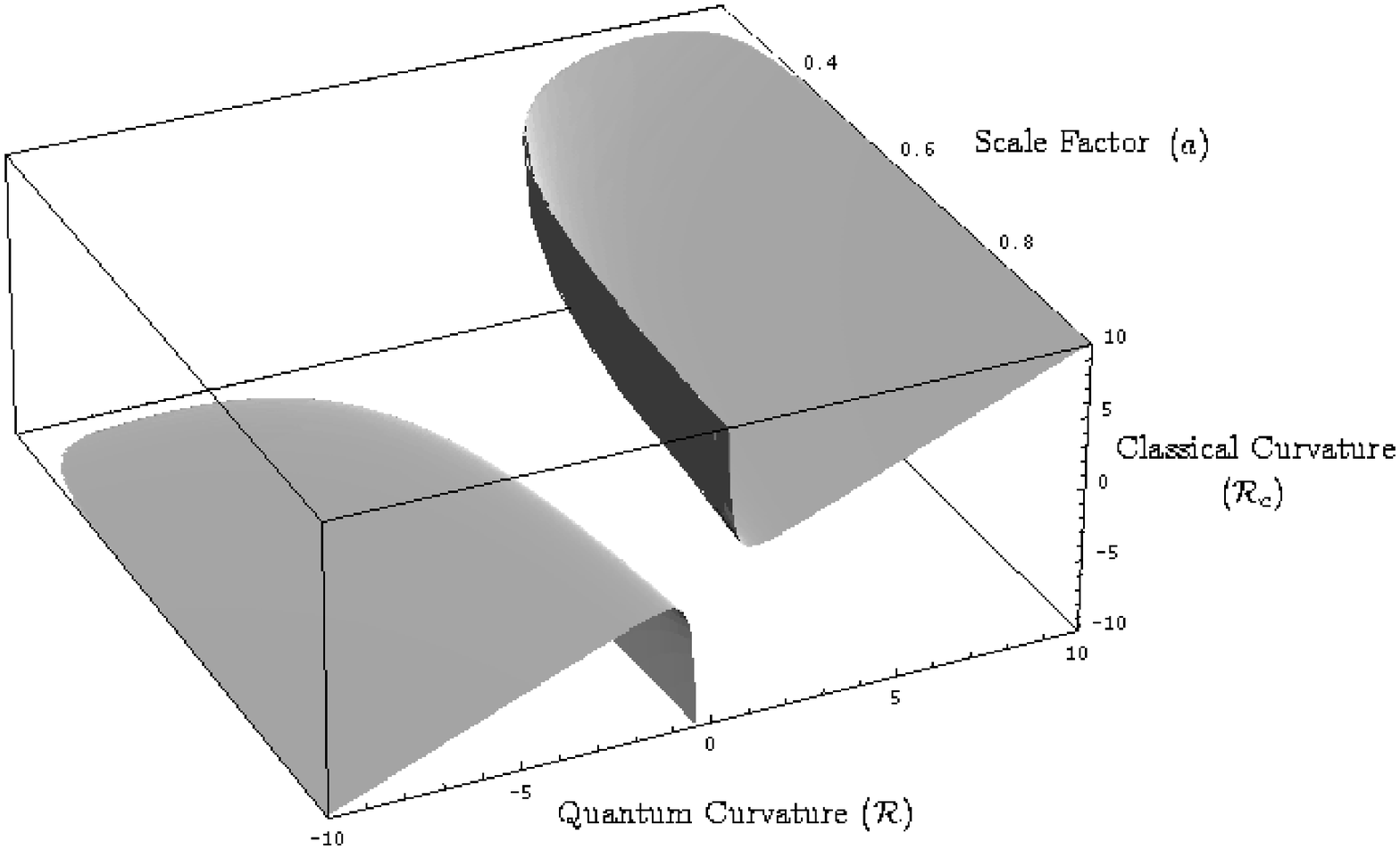}
\end{center}
\caption{Quantum curvature ($\r$) as a function of the scale factor ($a$) and classical curvature ($\r_c$) for the case $f(\r)=\r+\alpha\r^2$. $\alpha$ is chosen to be $\alpha^3=\hbar^2=\ell_p^4$.}
\label{fig2}
\end{figure}

\begin{figure}
\begin{center}
\epsffile{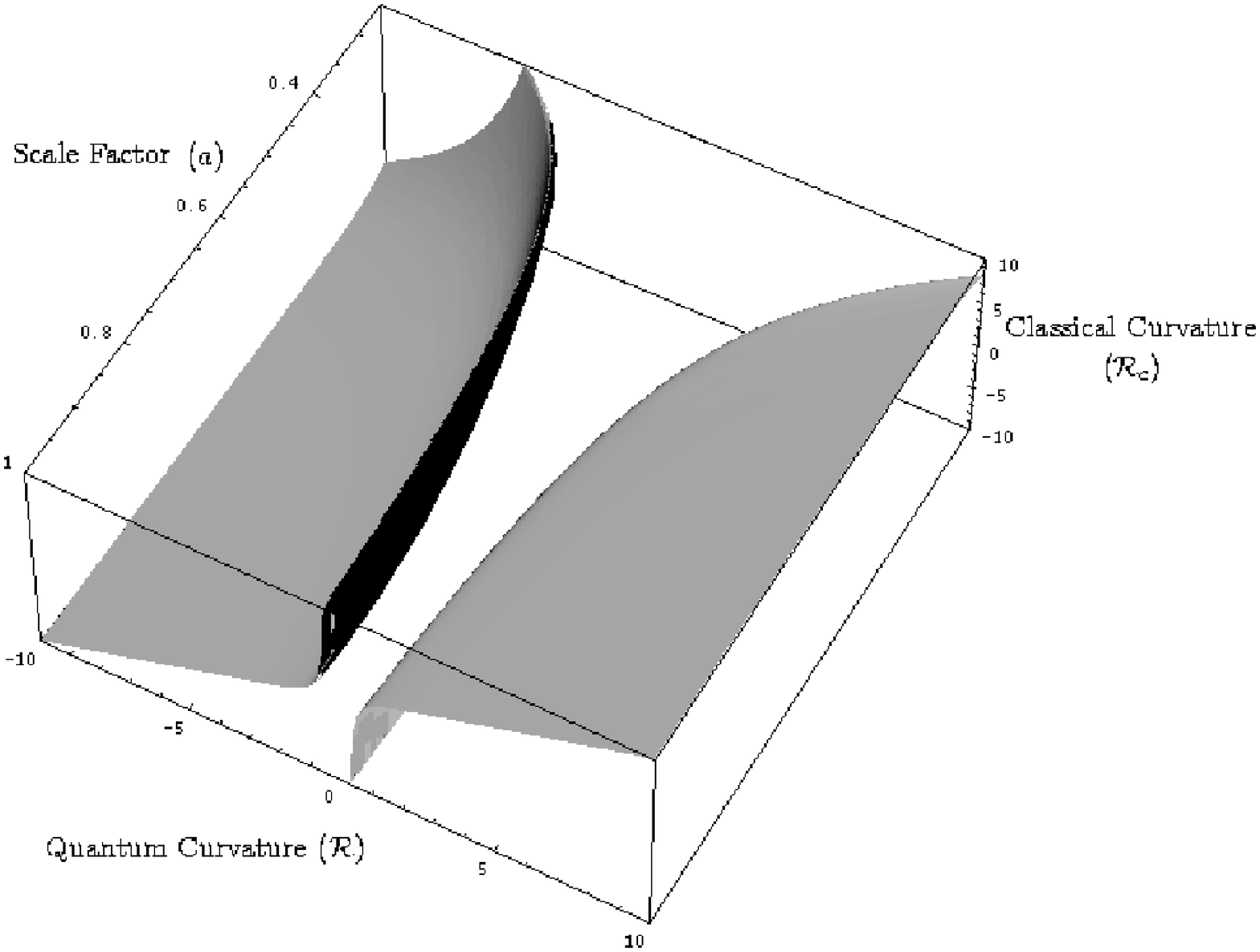}
\end{center}
\caption{Quantum curvature ($\r$) as a function of the scale factor ($a$) and classical curvature ($\r_c$) for the case $f(\r)=\r+\alpha\r^2$. $\alpha$ is chosen to be $\alpha^3=-\hbar^2=-\ell_p^4$.}
\label{fig3}
\end{figure}

As an example choosing $\beta\r_0 z=\alpha\r^2$ we get:
\begin{equation}
{\cal Q}=\frac{\hbar^2}{2\alpha^2a^3\r^2}(1-2\alpha\r)
\end{equation}
and
\begin{equation}
\r -\r_c=-\frac{\hbar^2}{2\alpha^3a^6\r^3}(1-\alpha\r)
\end{equation}
The behavior of the quantum potential and quantum curvature are plotted in figures (\ref{fig1}), (\ref{fig2}) and (\ref{fig3}). It has to be noted that this approximate solution is valid only when $|\r|<<1/|\alpha|$.

As it is clear from figures (\ref{fig1}), (\ref{fig2}) and (\ref{fig3}), four regions are distinguishable:
\begin{itemize}
\item The region $a$ large and $\r_c$ small in which the quantum potential goes to zero and and the quantum curvature goes to a flat surface on which quantum curvature equals to classical one. This is the classical regime.
\item The region $a$ and $\r_c$ large in which again the quantum potential goes to zero. Although this is a classical regime, the classical trajectories do not go this region (we have not a universe with large $a$ and large curvature).
\item The region $a$  and $\r_c$ large in which the quantum potential blows up. In this regime the quantum curvature is small. This is a highly quantum regime. Note that classical trajectories do not come to this region (we have not a universe with large $a$ and large curvature).
\item And finally the region $a$ small and $\r_c$ large in which quantum curvature has a nonzero finite value. Again this is a quantum region. In this regime quantum trajectories deviate from the classical ones as we have small scale factor but the curvature does not blow up. This shows that as it is discussed previously\cite{self}, the quantum potential corrections avoid the singularity of FRW model.
\end{itemize}

\epsfxsize=5in
\epsfysize=4.53in

\epsfxsize=5in
\epsfysize=3.73in

\epsfxsize=5in
\epsfysize=4.78in

This solution does not satisfy the criteria (\ref{sta}) for having de-Sitter stable solution, but in the limit $\r\rightarrow\infty$ and in the limit $a\rightarrow\infty$ (our present universe) we have  a  fixed point.

As a second example we choose $\beta \r_0 z=\alpha/\r$ leading to:
\begin{equation}
{\cal Q}=\frac{5\hbar^2\r^2}{4\alpha^2a^3}(\alpha+\r^2)
\end{equation}
and
\begin{equation}
\r -\r_c=-\frac{5\hbar^2\r^4}{4\alpha^3a^6}(\alpha+2\r^2)
\end{equation}

This solution does not satisfy the criteria (\ref{sta}) for having de-Sitter stable solution, but again in the limit $\r\rightarrow 0$ and in the limit $a\rightarrow\infty$ (our present universe) we have a fixed point.

\begin{figure}
\begin{center}
\epsffile{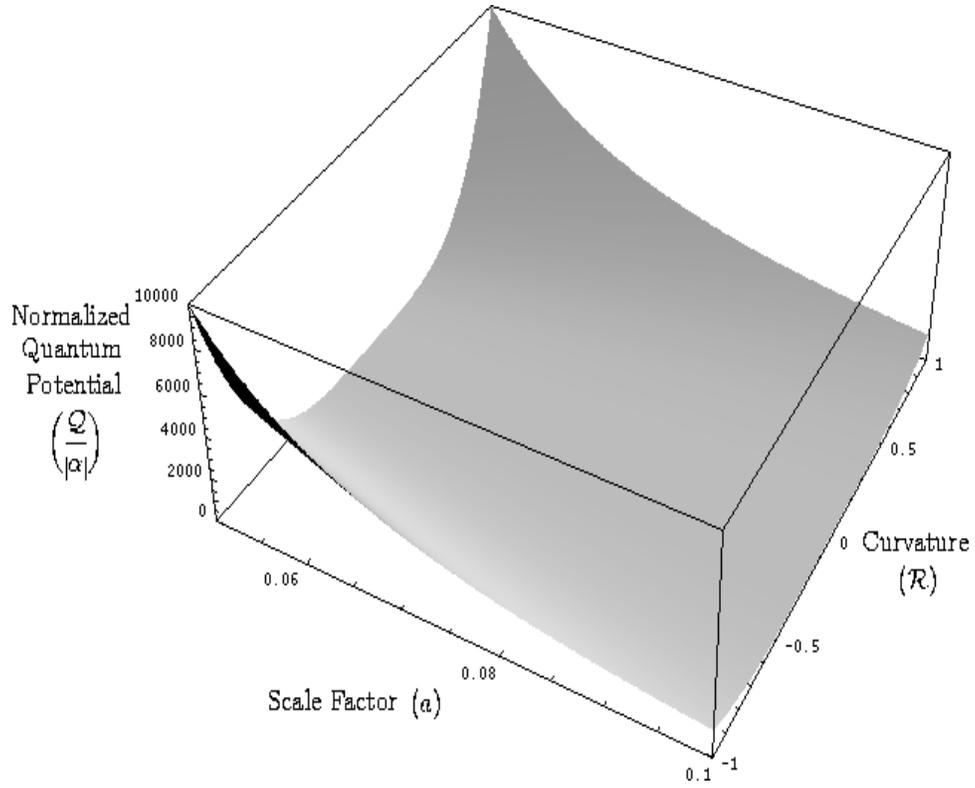}
\end{center}
\caption{Normalized quantum potential (${\cal Q}/|\alpha|$) as a function of the scale factor ($a$) and quantum curvature ($\r$) for the case $f(\r)=\r+\alpha/\r$. $\alpha$ is chosen to be $|\alpha|^3=\hbar^2=\ell_p^4$.}
\label{fig4}
\end{figure}

\begin{figure}
\begin{center}
\epsffile{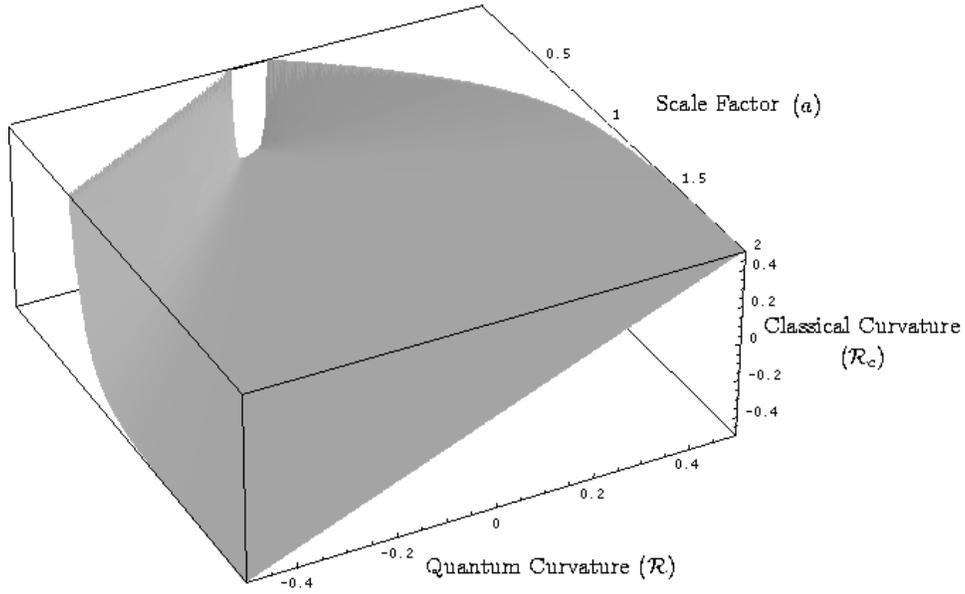}
\end{center}
\caption{Quantum curvature ($\r$) as a function of the scale factor ($a$) and classical curvature ($\r_c$) for the case $f(\r)=\r+\alpha/\r$. $\alpha$ is chosen to be $\alpha^3=\hbar^2=\ell_p^4$.}
\label{fig5}
\end{figure}

\begin{figure}
\begin{center}
\epsffile{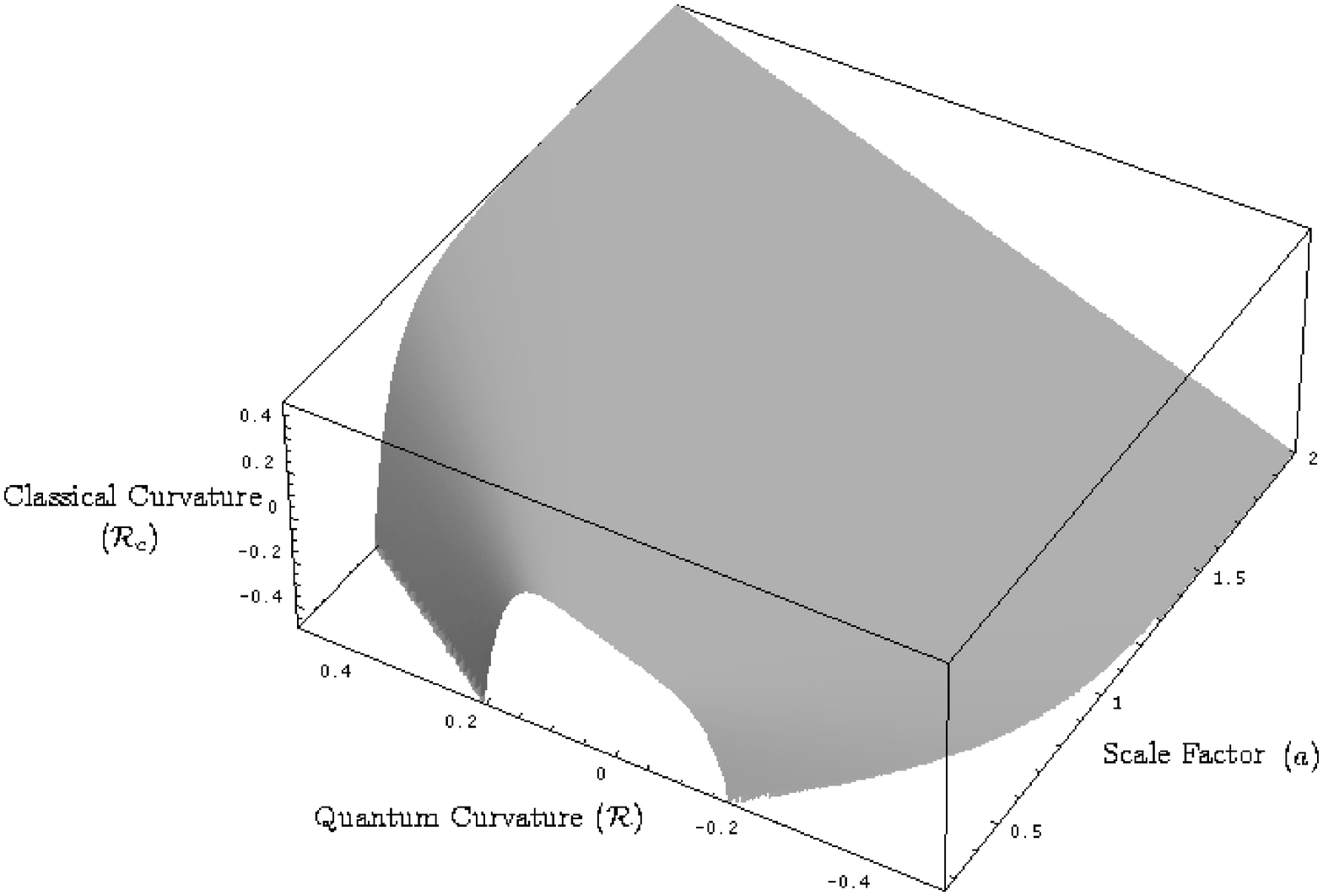}
\end{center}
\caption{Quantum curvature ($\r$) as a function of the scale factor ($a$) and classical curvature ($\r_c$) for the case $f(\r)=\r+\alpha/\r$. $\alpha$ is chosen to be $\alpha^3=-\hbar^2=-\ell_p^4$.}
\label{fig6}
\end{figure}

The behavior of the quantum potential and quantum curvature are plotted in figures (\ref{fig4}), (\ref{fig5}) and (\ref{fig6}).It has to be noted that this approximate solution is valid only when $|\r|>>\sqrt{|\alpha|}$.

From figures (\ref{fig4}), (\ref{fig5}) and (\ref{fig6}), one can distinguish three regions:
\begin{itemize}
\item The region $a$ large and $\r_c$ small in which the quantum potential goes to zero and and the quantum curvature goes to a flat surface on which quantum curvature equals to classical one. This is the classical regime.
\item The region $a$ and $\r_c$ large in which again the quantum potential is small. Although this is a classical regime, the classical trajectories do not go this region (we have not a universe with large $a$ and large curvature).
\item And Finally the region $a$  small and $\r_c$ large in which the quantum potential goes up. In this regime the quantum curvature is small. This is a highly quantum regime. Note that classical trajectories do not come to this region (we have not a universe with large $a$ and large curvature). In this regime quantum trajectories deviate from the classical ones as we have small scale factor but the curvature does not blow up. Again the quantum potential washes out the classical singularity. 
\end{itemize}

\epsfxsize=5in
\epsfysize=4.06in

\epsfxsize=5in
\epsfysize=3.92in

\epsfxsize=5in
\epsfysize=4.32in
\section{Conclusions}
Here we have quantized the metric $f(\r)$ theory of a flat cosmological model using the causal interpretation of quantum mechanics. It is shown that the acceleration of the universe depends both on the form of $f$ and on the quantum potential. An important result is that according to this Bohmian quantum picture the quantum curvature differs in form from the classical one by a term proportional to the quantum force as equation (\ref{qeq3}) shows. 

In order to investigate the behavior of such a model we considered two specific $f$'s ($f=\r+\alpha/\r$ and $f=\r+\alpha\r^2$), and obtained the dependence of the quantum gravity on the scale factor and the quantum curvature and also the dependence of quantum curvature on the scale factor and the classical curvature.

\textbf{Acknowledgment} This work is partly supported by a grant from university of Tehran and partly by a grant from center of excellence of department of physics in the structure of matter.

\end{document}